\documentclass{article} 
\usepackage{paper,times}

\usepackage{graphicx} 

\usepackage{xurl} 

\usepackage[breaklinks=true]{hyperref}
\usepackage{url}

\title{When Neural Implant meets Multimodal LLM: A Dual-Loop System for Neuromodulation and Naturalistic Neuralbehavioral Research}

\author{
Edward Hong Wang$^*$ \\ 
Harvard University \\
\And
Cynthia Xin Wen$^*$ \\
University of Sydney
}

\usepackage[symbol]{footmisc}

\begin{document}

\begin{center}
{\LARGE\bf When Neural Implant meets Multimodal LLM: A Dual-Loop System for Neuromodulation and Naturalistic Neuralbehavioral Research}

\vspace{0.5cm}

{\large
\begin{tabular}{ccc}
Edward Hong Wang$^*$ && Cynthia Xin Wen$^*$ \\
Harvard University && University of Sydney \\
\end{tabular}
}

\vspace{0.3cm}
\href{https://wxcynthia.github.io/AImonitor/}{\textcolor{blue}{https://wxcynthia.github.io/AImonitor/}}
\end{center}

\footnotetext[1]{These authors contributed equally to this work.}

\begin{abstract}
We propose a novel dual-loop system that synergistically combines responsive neurostimulation (RNS) implants with artificial intelligence-driven wearable devices for treating post-traumatic stress disorder (PTSD) and enabling naturalistic brain research. In PTSD Therapy Mode, an implanted closed-loop neural device monitors amygdala activity and provides on-demand stimulation upon detecting pathological theta oscillations, while an ensemble of wearables (smart glasses, smartwatches, smartphones) uses multimodal large language model (LLM) analysis of sensory data to detect environmental or physiological PTSD triggers and deliver timely audiovisual interventions. Logged events from both the neural and wearable loops are analyzed to personalize trigger detection and progressively transition patients to non-invasive interventions. In Neuroscience Research Mode, the same platform is adapted for real-world brain activity capture. Wearable-LLM systems recognize naturalistic events (social interactions, emotional situations, compulsive behaviors, decision making) and signal implanted RNS devices (via wireless triggers) to record synchronized intracranial data during these moments. This approach builds on recent advances in mobile intracranial EEG recording and closed-loop neuromodulation in humans \citep{BRAINInitiative2023} \citep{Mobbs2021}. We discuss how our interdisciplinary system could revolutionize PTSD therapy and cognitive neuroscience by enabling 24/7 monitoring, context-aware intervention, and rich data collection outside traditional labs. The vision is a future where AI-enhanced devices continuously collaborate with the human brain, offering therapeutic support and deep insights into neural function, with the resulting real-world context rich neural data, in turn, accelerating the development of more biologically-grounded and human-centric AI.

\end{abstract}

\section{Introduction}\label{s:introduction}
Post-traumatic stress disorder (PTSD) is a debilitating psychiatric condition characterized by intrusive memories, hyperarousal, and avoidance behaviors triggered by trauma-associated cues. While standard treatments like psychotherapy and medication can be effective, many patients experience breakthrough symptoms outside of clinical settings \citep{Sadeghi2022}. Episodes of intense fear or flashbacks often occur unpredictably in everyday environments, beyond the reach of therapists. This disconnect between clinic and real world hampers timely intervention and limits our understanding of PTSD's neural underpinnings in natural contexts \citep{Sadeghi2022}. Recent technological advances suggest a path forward: wearable sensors, mobile computing, and implanted neuro devices now allow continuous monitoring of physiology and brain activity in freely moving humans \citep{BRAINInitiative2023}. In parallel, artificial intelligence – especially large language models (LLMs) capable of multimodal data analysis – offers new possibilities for interpreting complex streams of data and guiding therapeutic responses in real time \citep{AlSaad2024}. Additionally, recent work on specialized AI-centric languages points to even more efficient, unambiguous communication within AI-driven systems, potentially enhancing our platform's ability to unify neural and contextual data \citep{Wang2025}.

This paper presents a novel system integrating responsive neural stimulation implants with an LLM-powered wearable platform to create a dual-loop closed-loop intervention for PTSD. The first loop is a Responsive Neurostimulation (RNS) Implant Loop, which provides immediate neuromodulation in the brain's emotional circuits when pathological activity is detected. The second loop is a Wearable–LLM Loop, which monitors the patient's environment and physiology to preemptively identify PTSD triggers and deliver calming sensory cues. By coordinating these two loops, the system can respond to PTSD symptoms both internally (at the neural circuit level) and externally (via sensory/contextual intervention), creating a comprehensive therapeutic shield around the patient. Logged data from both loops are continuously fed into a multimodal AI model to refine trigger detection and personalize therapy over time. We hypothesize that this dual-loop approach can not only improve symptom control in severe PTSD but also facilitate a gradual transition toward non-invasive management: as the wearable system 'learns' and becomes increasingly adept at preemptively managing triggers, the reliance on implant-based interventions can be reduced or ultimately eliminated, offering patients the potential to maintain symptom control without intracranial neural stimulation.

Beyond clinical therapy, our system doubles as a powerful tool for naturalistic neuroscience research. By leveraging wearable sensors to detect specific real-life events and triggering high-resolution brain recordings at those moments, researchers can collect context-tagged neural data during unconstrained daily activities. This capability addresses a long-standing challenge in neuroscience: studying brain activity underlying complex behaviors that cannot be reproduced in lab experiments \citep{Mobbs2021}. Our approach draws inspiration from recent studies that successfully recorded deep brain signals in ambulatory human subjects using wearable EEG/implant setups \citep{BRAINInitiative2023}. We extend this paradigm by incorporating Multimodal LLM to recognize nuanced social or behavioral events (e.g. engaging in conversation, experiencing spontaneous joy or stress) and time-lock them with intracranial recordings. The result is a richly annotated brain dataset collected in the wild, opening new frontiers in understanding neural mechanisms of emotion, memory, and behavior in real-world contexts.

In the following sections, we first review relevant background: emerging closed-loop neurostimulation therapies for PTSD and the state-of-the-art in wearable sensing and AI intervention for mental health. We then detail the architecture of the dual-loop system, describing its two modes (therapy and research) and their implementation. We anchor our approach in existing evidence, including a recent first-of-its-kind closed-loop amygdala stimulation trial in PTSD \citep{Gill2023} and pioneering mobile brain-recording experiments by Suthana and colleagues \citep{SuthanaLab2025}. Finally, we discuss the broader vision: how AI-powered multimodal wearable systems could revolutionize neuroscience and psychiatry by enabling continuous, context-aware brain monitoring and intervention, transcending the confines of laboratory and clinic. We emphasize the mutual benefits for both AI (which gains grounding in human neural data) and neuroscience (which gains adaptive analytical tools and unprecedented data access), and call for an interdisciplinary effort to realize this vision.

\section{Background and Rationale}\label{s:background}

\subsection{Closed-Loop Neurostimulation for PTSD}\label{ss:neurostimulation}
PTSD has well-defined neural correlates, particularly in limbic regions such as the amygdala and hippocampus which govern fear processing and memory. Aberrant oscillatory activity in these circuits is associated with PTSD symptoms like hypervigilance and flashbacks. However, until recently, directly recording and modulating these deep brain signals in patients with PTSD was uncharted territory. In a groundbreaking pilot clinical trial, Gill et al. (2023) implanted depth electrodes in the amygdala of two patients with treatment-resistant PTSD to record intracranial electroencephalography (iEEG) continuously over a year \citep{Gill2023}. During exposure to trauma-related stimuli and periods of natural symptom exacerbation, elevated amygdala theta-band (5–9 Hz) activity was identified as a consistent neural signature of distress \citep{Gill2023}. This biomarker was then used to trigger responsive neurostimulation: whenever the amygdala's low-frequency power surged past a threshold, an electrical stimulation was delivered automatically. The results were promising – closed-loop stimulation led to significant reductions in PTSD symptom severity over one year, along with dampening of the previously excessive theta activity \citep{Gill2023}. These findings provide human proof-of-concept that a closed-loop device can detect pathological brain states in PTSD and intervene in real time to alleviate symptoms \citep{Gill2023}. Notably, they highlight amygdala theta oscillations as a viable control signal or "trigger" for device-driven therapy.

The device used in such trials is analogous to the Responsive Neurostimulation (RNS) system, originally developed for epilepsy. The NeuroPace RNS™ is a fully implanted neurostimulator that continually monitors brain EEG from implanted leads and delivers electrical pulses when detecting epileptiform activity. It consists of a cranial neurostimulator connected to one or more electrodes (depth leads targeting deep structures and/or cortical strip leads)

\begin{figure}[h]
\includegraphics[width=\textwidth]{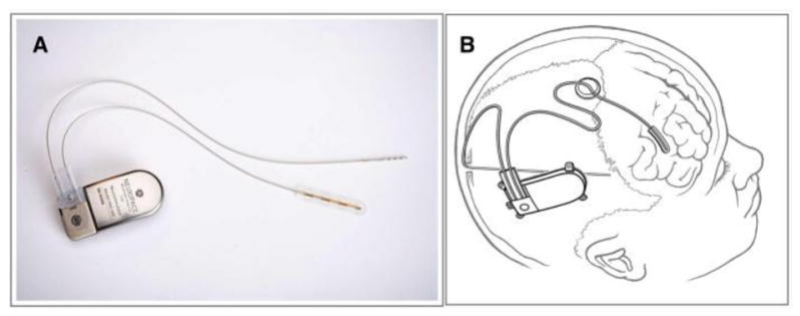}
\caption{The responsive neurostimulation device (RNS). (A) The RNS implant with a four-contact depth lead and a four-contact cortical strip. (B) Artistic schematic of the RNS implanted in the skull with a depth lead into a deep brain target (e.g., amygdala) and a strip electrode on cortical surface. \citep{Englot2017}}
\label{fig:rns}
\end{figure}

This technology, FDA-approved for refractory epilepsy, is now being repurposed in experimental therapies for psychiatric disorders. The PTSD pilot by Gill et al. (2023) likely utilized a similar RNS approach – detecting a neural signature (amygdala theta burst) and triggering stimulation to disrupt the pathological state. From epilepsy we know that RNS can operate continuously and store electrographic logs of detections and stimulations over long periods in patients' everyday lives. Importantly, such neural implants have been proven safe and tolerable in patients, with the hardware fully implanted under the scalp and requiring only periodic clinical downloads (every 1-2 days) and battery charging (typically lasting 4-8 years). This makes chronic closed-loop neuromodulation feasible for disorders like PTSD that manifest outside hospital settings.

Beyond PTSD, closed-loop deep brain stimulation (DBS) is being actively explored in other neuropsychiatric conditions, lending credibility to the concept. For example, in obsessive-compulsive disorder (OCD), recent work by Nho et al. (2023) identified electrophysiological biomarkers of obsessive thought patterns in the ventral striatum and used them to drive responsive DBS, resulting in durable reductions in compulsive behaviors \citep{Nho2023}. Similarly, there are case reports of dual-use RNS devices aiding both epilepsy and comorbid psychiatric symptoms \citep{Kellogg2024}. These advances echo a broader trend: closed-loop neuromodulation – once confined to research labs – is moving into clinical reality for treating circuit-based disorders. PTSD, with its discrete triggerable episodes and identifiable neural rhythms \citep{Gill2023}, is a prime candidate for such technology.

Nevertheless, neural implants alone address only part of the PTSD puzzle. They act after the brain has already generated a pathological pattern (e.g. a flashback onset) and modulate neural activity internally. They do not directly engage with the patient's external environment or conscious experience of triggers. Moreover, implants like RNS, while effective, are invasive and expensive; ideally, one would taper their use over time if symptoms can be managed non-invasively. This motivates pairing the RNS loop with an external loop that can detect and mitigate triggers in the sensory realm, working towards eventually offloading the implant.

\subsection{Wearable Sensors and AI for Trigger Detection and Intervention}\label{ss:wearables}
Wearable devices – such as smartwatches, fitness bands, and smart glasses – have proliferated in recent years and are increasingly used for mental health monitoring. These devices continuously measure physiological signals (heart rate, galvanic skin response, movement, etc.) and can also capture aspects of the environment (sound via microphones, imagery via outward-facing cameras, location via GPS). Researchers are leveraging wearables to recognize stress and anxiety states in the wild, a concept often termed "digital phenotyping" of psychiatric conditions \citep{Kalisperakis2023}. For PTSD specifically, studies have shown it is feasible to detect moments of hyperarousal or distress from wearable sensor patterns. For instance, Sadeghi et al. (2021) collected days of smartwatch data (heart rate and accelerometry) from 99 veterans with PTSD and applied machine learning to identify windows corresponding to self-reported PTSD episodes \citep{Sadeghi2022}. Their best model (an XGBoost classifier) detected the onset of PTSD hyperarousal events with over 83\% accuracy (AUC 0.70) \citep{Sadeghi2022}. Notably, features like elevated average heart rate and sudden accelerations (startle-like movements) were most predictive of episodes \citep{Sadeghi2022}. This demonstrates that physiologic precursors of PTSD episodes can be picked up by commercial wearables, enabling continuous monitoring and early warning outside the clinic. The authors emphasize that such systems could "address a vital gap in just-in-time interventions for PTSD self-management outside of scheduled appointments" \citep{Sadeghi2022} – essentially, empowering real-time support in everyday life rather than relying solely on periodic therapy sessions.

Other work supports the feasibility of wearable-triggered interventions. A notable example is the Hero Trak system, a PTSD monitoring platform tested with veterans during group cycling events \citep{TexasAMEngineering2017}. In field trials, Hero Trak used off-the-shelf smartwatches (Moto 360) to continuously track physiology and employed algorithms to flag possible PTSD trigger events \citep{TexasAMEngineering2017}. Participants and developers reported that the system successfully identified arousal spikes during natural activities and that users valued the ability to "track, identify, and deal with PTSD triggers" in real time \citep{TexasAMEngineering2017}. This underscores that patients are receptive to wearable technology aiding their coping in real-world scenarios. Similarly, the FDA-approved NightWare system provides a closed-loop therapy for PTSD-related nightmares using just a smartwatch and phone. NightWare monitors heart rate and motion during sleep to detect nightmares (characterized by a sharp rise in heart rate and movement) and upon detection, it triggers the smartwatch to vibrate, subtly arousing the sleeper from the nightmare without fully waking them \citep{NightWare}. Over repeated use, NightWare's AI algorithms personalize the detection and intervention parameters to the individual's sleep patterns \citep{NightWare}. This is a compelling proof-of-concept that even without an implant, wearables can implement a closed-loop therapeutic feedback: sensing a physiologic marker of distress and delivering a somatic intervention (vibration cue) to improve outcomes (in this case, sleep quality).

The next frontier is combining such multimodal sensor inputs with advanced AI – specifically, multimodal large language models (LLMs) – to interpret complex contexts and decide tailored interventions. Traditional algorithms (like threshold-based triggers or classical classifiers) have shown promise in detecting PTSD-related events, but they are often limited to specific signal types (e.g., heart rate) and may struggle with the rich, contextual nature of triggers (e.g., a particular visual cue or phrase that reminds a patient of trauma). Modern LLMs such as GPT-4o are inherently adept at integrating information across different modalities and semantic levels, especially when extended to process images, audio, and time-series data in addition to text \citep{AlSaad2024}. Recent reviews highlight that multimodal LLMs in healthcare can synthesize sensor data with knowledge to provide decision support beyond what single-signal models can do \citep{AlSaad2024}. In our context, an Multimodal LLM-based system could, for example, take as input a live stream of a patient's vitals, a live video stream from smart glasses, and infer whether the patient is encountering a known trigger situation – essentially performing Multimodal LLM-driven situational awareness. If the model recognizes a scenario likely to provoke PTSD symptoms (for instance, the patient is back in a crowded setting similar to their trauma environment, or a loud bang resembling gunfire occurred), it can initiate preventative calming measures. The advantage of an LLM here is the ability to incorporate semantic understanding (e.g., recognizing the scenario "being in a crowded marketplace" from audio-visual cues, or understanding that a sudden loud sound plus the patient's startled bio marker reaction such as heart rate rush equals a high likelihood of trigger). This goes beyond simple pattern matching and moves into AI that understands context, an area where pre-trained foundation models excel by virtue of having absorbed world knowledge.

Several enabling technologies for such a system are already in place. Smart glasses with integrated cameras, heads-up displays, mic and speaker are commercially available and can stream the user's video and deliver audiovisual feedback. Smartwatches capture heart rate, variability, temperature, and motion. Cloud-connected smartphones can serve as the computational and data transmission hub, running Multimodal LLM on device or in cloud to fuse these data. Augmented Reality exposure therapy research is starting to use AI to inject realistic social cues into patients' environments for therapeutic purposes \citep{Javanbakht2024}, indicating that the concept of AI-mediated context in PTSD treatment is gaining traction. Our approach leverages these components but in a unique closed-loop configuration that links directly with neural activity via the implant.

In summary, wearable sensors and AI algorithms can increasingly detect PTSD triggers or early warning signs in natural settings, and even administer basic interventions (app alerts, Audio or Visual content). What has been missing is a way to integrate these external-loop interventions with the internal-loop neural interventions, so that both the mind and brain are supported in tandem. The following section outlines our dual-loop system that bridges this gap.

\section{Dual-Loop System for PTSD Therapy}\label{s:dualloop}
Our proposed system operates in a PTSD Therapy Mode that involves two synergistic feedback loops (Figure 2). The first is the RNS Implant Loop, focused on real-time neural monitoring and stimulation, and the second is the Wearable–LLM Loop, focused on peripheral monitoring and sensory intervention. Both loops function continuously and concurrently, sharing data to improve overall performance. Below we describe each loop in detail and then their integration for personalized therapy.

\subsection{RNS Implant Loop (Neurofeedback Loop)}\label{ss:rnsloop}
This loop centers on an implanted responsive neurostimulator placed in the patient's skull (e.g., a NeuroPace RNS or similar closed-loop DBS device). Leads from the device target key brain regions implicated in PTSD – in our prototype, bilateral amygdala depth electrodes are used, given the amygdala's crucial role in fear and emotional memory. The implant continuously records local field potentials (LFPs) from the amygdala (and potentially connected regions like hippocampus or ventromedial prefrontal cortex if additional electrodes are placed) and analyzes the signals in real time on-device. It is configured to detect abnormal patterns associated with PTSD episodes, most notably the excessive theta-band power that was identified as a biomarker in prior research \citep{Gill2023}. The detection algorithm can use power spectral thresholds or a machine-learning classifier trained on the patient's own baseline recordings. When the algorithm detects a likely prodrome of PTSD symptoms – for example, a sudden surge in 5–9 Hz oscillatory power exceeding a predefined threshold for a certain duration – the device automatically delivers a brief electrical stimulation to the amygdala. The stimulation parameters (current, pulse width, frequency) would have been titrated previously to ensure they disrupt the pathologic activity without causing adverse effects \citep{Gill2023}. The goal of the stimulation is to abort the developing neural cascade of a PTSD flashback or panic response, nipping it in the bud at the circuit level. Because this happens on the order of seconds and entirely inside the brain, it provides an immediate safety net even if the external trigger is not yet addressed.

Crucially, every detection and stimulation event is time-stamped and logged in the device memory. The neural data around those events can be stored as snippets of iEEG. These logs include whether the stimulation succeeded in quelling the abnormal activity (e.g., did theta power subside after the pulse?). Over time, this creates a personalized record of the patient's neural "storms" and the efficacy of interventions. The implant loop thus provides an objective neural measure of PTSD episodes, which is valuable for both therapy and research.

It is worth noting that responsive stimulation in the amygdala for PTSD is still experimental, but early evidence suggests it can be done safely and effectively \citep{Gill2023}. Amygdala-focused DBS has been tested in animals for fear conditioning, and the pilot human trial reported no serious adverse effects related to stimulation \citep{Gill2023}. Our system would involve an implanted device very similar to existing RNS for epilepsy, whose safety profile is well established (over 3,000 patient implants with many years of cumulative use). The novelty lies in programming its detection algorithm for a PTSD-specific signal and using it in tandem with wearables.

\subsection{Wearable–LLM Loop (Peripheral Feedback Loop)}\label{ss:wearableloop}
In parallel to the neural loop, the patient wears a suite of unobtrusive sensors that continuously monitor their physiological state and surroundings. In our design, this includes: a smartwatch for heart rate (including variability), acceleration (motion, tremor), skin temperature, and galvanic skin response (stress-related sweating); smart glasses with an embedded camera and AR display, a microphone and a speaker to stream video feed and provide audiovisual feedback; and the patient's smartphone which serves as a central hub and additional sensor (GPS location, time of day, perhaps additional processors). Data from these sensors stream in real-time to an on-device or in cloud Multimodal LLM, tailored for interpreting the combined data. This model has been efficiently prompted and fine-tuned with psychiatric contextual data so that it can perform tasks such as: detecting if the patient is currently in a potentially triggering environment, predicting the patient's emotional state from biometrics and behavior, and deciding if/what intervention is warranted.

The LLM's analysis happens continuously but most of the time it remains in a passive monitoring mode. When certain conditions are met – either gradually building stress signs or a sudden trigger – the system transitions to intervention. Detection in the wearable loop might work as follows: the LLM looks for convergent cues of distress. For example, a rapid heart rate increase (beyond what would be expected from activity, as the accelerometer can contextualize), a change in voice tone/breathing and visual input that matches known trigger contexts (e.g., the camera sees a scene that resembles the location of the patient's trauma, or identifies a person or object the patient associates with trauma). One can imagine the LLM parsing these inputs almost like a human would: "The patient's heart rate jumped from 80 to 130 in a few seconds without heavy physical activity; the audio picked up a loud bang that sounds like a car backfire; the patient is now in a crowded street (per GPS and imagery) – this combination suggests the patient might be re-experiencing a combat memory." At this point, the wearable loop flags a likely PTSD trigger event.

Intervention in the wearable loop is then deployed through two main channels: visual and auditory. Via the smart glasses' display, the system can alter the patient's visual experience. In a simple form, it could dim or block certain visual stimuli – for instance, darkening the periphery of the visual field to reduce overwhelming inputs, or overlaying a calming image (such as a safe location or a loved one's picture) as a grounding cue. More dynamically, it might project a subtle augmented reality scene that promotes calm (imagine gentle waves or a soothing color tone washing over the real scene). Concurrently, via speakers in the smart glasses, the system provides auditory comfort: this could be a pre-recorded message from the patient's therapist (“Remember to breathe, you are safe here"), a brief mindfulness instruction generated by the LLM, or even just calming nature sounds or music tailored to the patient's preferences. The LLM can generate or select the spoken content based on context – for example, if the trigger was related to combat, the message could be specifically reassuring about being in the present, whereas if it was social anxiety, the message might encourage the patient to engage in a coping skill.

These interventions are informed by established PTSD grounding techniques, essentially delivered through an automated AI assistant. The wearable loop thus tries to disrupt the trigger before it fully captures the patient's mind, providing a cognitive and sensory anchor to the present safety. Importantly, the wearable interventions are designed to be as un-intrusive as possible to avoid embarrassment or drawing attention in public; AR visualizations can be private to the glasses, and audio can be near-silent to others. The smartphone can also vibrate gently as an additional cue if needed.

Like the RNS device, the wearable loop logs its decisions and outcomes. It notes the sensor readings that led to each trigger detection, the type of intervention given, and ideally the patient's response (did the patient's heart rate come down? Did they resume normal activity?). The LLM can summarize these events in a daily log, which can be reviewed by clinicians to adjust therapy or by the LLM itself to conduct in context learning.

\subsection{Integration and Personalization}\label{ss:integration}
The true power of the system emerges when combining the two loops. While each can operate independently, they inform each other to create a comprehensive closed-loop across the brain and body. Consider a scenario: The patient is walking down a street when a triggering event occurs (say a car backfire reminiscent of gunfire). The wearable loop may detect the context and start a calming intervention at nearly the same time the RNS loop detects surging amygdala theta and delivers a neurostimulation. In effect, one loop is working from "outside-in" (reducing the external trigger impact) and the other from "inside-out" (directly damping the internal fear network). The patient might briefly feel the heart skip and a jolt of anxiety, but then immediately a sense of calm as the brain stimulation and AR/audio cues take effect in concert. Ideally, the episode is aborted – the patient does not spiral into a panic attack or dissociative flashback.

Now consider the data collected: the RNS log would show a detection at timestamp T, and a stimulation delivered. The wearable log shows an audio event (loud bang) at T minus 1 second, a heart rate spike at T, and interventions (glasses dimming, soothing voice) at T plus 2 seconds. By aligning these logs, the system (and clinicians) can see a holistic picture of the event: what external trigger precipitated it, how the body responded, what the brain activity was like, and how effective each intervention was. Over many such events, patterns emerge. Perhaps the logs reveal that the RNS device often triggers just after a certain type of external cue. The LLM can learn to anticipate those by recognizing them faster and even preemptively engaging the patient in grounding techniques (for example, if it recognizes they just entered a location that has triggered past episodes, it might proactively play a brief calming exercise). Conversely, the logs might show some events where the wearable loop missed a trigger that the implant caught (e.g., a panic attack with an internal trigger like a memory, not an external cue). For those, the system could learn the subtle wearable signs that accompany the internal trigger (maybe a certain breathing pattern or skin temperature) and improve detection next time. This cross-talk essentially allows the neural data to supervise the AI model training and the AI's contextual data to refine the neural device's trigger criteria.

Personalization is further achieved by involving the patient and clinician in the loop. The patient could rate their subjective distress after each logged event via a phone interface, which provides ground truth for the AI's assessments. The clinician can adjust parameters – e.g., raise or lower the sensitivity of the wearable trigger detection if it's too often false-alarming or missing events. In context reinforcement learning human feedback (RLHF) could be applied to optimize the Multimodal LLM continuously. Over time, one might find that for this patient, certain trigger types are almost entirely handled by the wearable loop interventions without needing the RNS to fire. In such cases, the RNS threshold might be made less sensitive (to only capture severe events) as the patient gains confidence that non-invasive strategies suffice for most triggers. In an ideal endpoint, the patient's PTSD could enter remission to the degree that the RNS seldom activates; at that stage, one could consider explanting the device or turning it off, allowing the patient to rely on the non-invasive wearable system alone. Even if the device remains implanted as a safety backup, the goal is that the patient experiences therapy as largely external and under their control, which may be psychologically empowering. This addresses the long-term vision of transitioning patients toward non-invasive treatment, using the implant as a bridge to get there.

From a technical perspective, integrating the loops requires secure communication. The wearable hub (smartphone) can be paired with the implant's controller. Modern RNS devices have wireless capabilities (typically magnetically activated telemetry, i.e. NFC like tapping a credit card); for real-time integration, we envision using a low-power wireless protocol (e.g., Bluetooth Low Energy) or build a tiny NFC chip into a wearable daily hat to allow the phone to poll the RNS for events periodically or be notified of detections. Data privacy and security are paramount given the sensitive nature of both neural and audiovisual data; all data can be processed locally on the patient's devices with cloud sync optional for clinician review under consent.

In summary, the dual-loop system in PTSD Therapy Mode creates a closed-loop ecosystem around the patient: the inner loop provides direct neurophysiological stabilization, and the outer loop provides situational awareness and guided coping. By sharing information, they enhance each other's ability to protect the patient from harm and facilitate learning and adaptation of the therapy to the individual's life.

\begin{figure}[!htb]
\includegraphics[width=\textwidth,page=2]{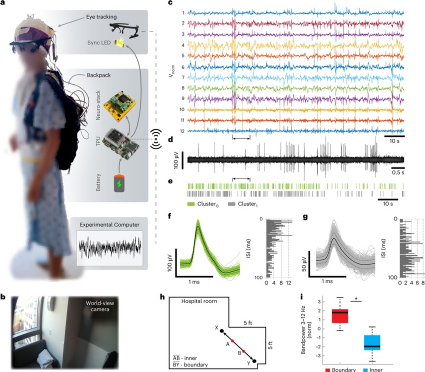}
\caption{Conceptual diagram of the dual-loop PTSD therapy system. The patient wears an RNS implant (brain icon) that monitors amygdala activity and stimulates when detecting a high-theta event. Simultaneously, wearable smart glasses and smart watch feed data to an on-device or in-cloud Multimodal LLM that detects triggers (e.g., a loud noise, elevated heart rate) and initiates visual and auditory feedback via smart glasses and smart phone. Both loops log events to a secure database. The LLM can also correlate implant activations with environmental triggers, conduct in context learning and refining its detection algorithms over time. Through this synergy, the system intervenes at both the brain level and experiential level to interrupt PTSD episodes.}
\label{fig:dualloop}
\end{figure}

\section{Naturalistic Neuroscience Research Mode}\label{s:researchmode}
In addition to its therapeutic application, our platform serves as a research tool for neuroscience, enabling the study of human brain activity during real-world experiences with unprecedented resolution and context. The system can be switched to a Research Mode in patients or research participants who have neural implants (either for PTSD as above, or for other clinical indications like epilepsy or OCD, or even exploratory research implants). In this mode, the primary aim is data collection: capturing synchronized streams of brain signals and contextual sensor data during natural behaviors and events. We outline how the components of the system are utilized for research and provide example use-cases.

\subsection{Context-Triggered Neural Recording in Everyday Settings}\label{ss:contextrecording}
Traditionally, human neuroscience experiments are done in laboratories or hospitals where conditions can be controlled. However, this control comes at the cost of ecological validity – the behaviors observed are mere proxies for the richness of daily life. A core vision of our approach is to take neuroscience into the real world, aligning with calls for a "mobile cognition" paradigm \citep{Mobbs2021} and "human computational ethology" \citep{Mobbs2021}. Our wearable-LLM system acts as a vigilant observer of the participant's environment and actions. Using similar multimodal LLM analysis as described for trigger detection, the AI can be configured to detect events that the researcher is interested in. These could be specific activities (e.g., the participant is playing a sport, or navigating through a new city), social interactions (e.g., the participant is engaged in conversation or a group setting), or internal states (e.g., signs of stress or compulsive eating in an addiction study). The key is that the AI can use high-level descriptors – it does not require rigid predefined triggers but can be flexibly trained or prompted to recognize complex scenarios (for example, "record data whenever the participant appears to laugh or whenever they exhibit a compulsive behavior like repeated hand-washing").

When such an event is detected, the wearable system can send a signal to the implanted device to initiate a recording. Modern neural implants (RNS, Medtronic Percept PC, etc.) often have the capability to store raw electrophysiological data for a short duration when triggered, primarily used in clinical contexts to save abnormal EEG segments or neural activity around detected events. In research mode, we exploit this feature deliberately: we essentially force-trigger the implant to record even if no "abnormal" neural event is happening, because we are interested in the neural correlates of the context identified by the AI. Practically, this could be achieved in multiple ways. One approach is using the standard patient controller for RNS devices: typically, patients can mark events by placing a magnet over the device (e.g., if they sense a seizure aura). Our system's wearable component (like a "smart hat") could include a small electromagnet or NFC (near-field communication) device that acts like a magnet placement to trigger the RNS. A more elegant approach is if the implant API allows direct wireless command to start recording; some research-oriented implants provide researcher access to trigger data capture via a paired tablet. In either case, upon AI detection of the target event, a message (or magnet pulse) is sent to the implant, which then saves, say, the previous and next 60 seconds of high-fidelity neural data from all channels.

Simultaneously, the wearable devices record the external data around that event – video from the glasses, physiological metrics from smart watches, geological data from smart phones – and time-stamp them. This yields a synchronized dataset: neural activity from inside the brain aligned with rich annotations of what the person was doing and experiencing. Such data is the holy grail for understanding brain-behavior relationships in an uncontrolled setting. It effectively realizes what Suthana et al envisioned: integrating intracranial recordings with peripheral measurements to study brain function in real-world contexts \citep{SuthanaLab2025}. Suthana's laboratory has already demonstrated components of this: in a study with epilepsy patients, they used a backpack-mounted Neuro-stack device and an eye-tracking headset to record single-neuron activity while patients freely ambulated and looked around a room \citep{Stangl2023}. They correlated neural oscillations and spiking with spatial behavior (e.g., approaching boundaries of the room) \citep{Stangl2023}.

\begin{figure}[h]
\caption{A participant in Suthana et al. (2023) wearing a backpack with the Neuro-stack (for iEEG recording) and an eye-tracking device. This enabled recording of hippocampal and cingulate neurons as the participant walked naturally in a room \citep{Stangl2023}. Such wearable setups mark the beginning of real-world neuroscience, capturing data during free behavior that was previously accessible only in animals \citep{DailyBruin2023}.}
\label{fig:neurostack}
\end{figure}

Our system builds on these advances by adding AI-driven context detection and the ability to trigger recordings during specific complex events, not just continuous recording during simple tasks. This is akin to having a research assistant following the participant around and shouting "mark now!" whenever an interesting behavior happens, but done automatically and objectively.

\subsection{Example Applications}\label{ss:applications}
\begin{itemize}
    \item \textbf{Social Neuroscience:} Human social interactions are difficult to simulate in labs. With our system, one could study how the hippocampus or amygdala behaves during genuine social encounters. For instance, the AI could be set to recognize when the participant is in a family gathering (via multiple faces detected in view, voices of known family members, etc.). When a warm emotional moment occurs (laughter, hugging), the system triggers a recording. Over many such natural interactions, researchers can analyze how deep brain signals (perhaps theta oscillations or single-unit firing in the amygdala) track affiliative emotions, attachment, or social anxiety in real life. This extends recent efforts in computational ethology to link natural behaviors with neural data \citep{Mobbs2021}. We might discover, for example, that the amygdala of a PTSD patient actually shows a different pattern in a happy social situation versus a stressful one, which could inform emotion-specific interventions.
    \item \textbf{Sports and Navigation:} If a participant with epilepsy has electrodes in motor or navigation-related areas, the system could trigger recordings during sports activities or free navigation in a city. Imagine analyzing motor cortex LFPs while a person plays a casual game of basketball in their driveway, using the wearable detection of "shooting a hoop" events to capture neural snapshots. This could provide insights into natural motor coordination or error-related neural signals when they miss a shot, for example. Similarly, recording hippocampal activity while a person naturally navigates a grocery store (detected by context cues) can complement virtual navigation studies and test if spatial memory circuits behave similarly in the wild \citep{Mobbs2021}.
    \item \textbf{Psychiatric Behavior Analysis:} For conditions like OCD or Tourette's syndrome, in patients who have deep brain electrodes for experimental therapy, the wearable AI loop can be trained to spot compulsive behaviors or tics. For OCD, say a patient has a compulsion of hand-washing or stress eating. The AI could learn the visual and motion signature of these rituals (e.g., repetitive hand movements at a sink) and trigger brain recordings during those behaviors at home. This would let researchers examine the closed-loop neural dynamics of compulsions: what signals in the nucleus accumbens or orbitofrontal cortex ramp up before a compulsion, and how they settle after completion. Such data could validate biomarkers for use in future closed-loop stimulation (similar to what Nho et al. identified in ventral striatum \citep{Nho2023}). It also helps assess if the implant's stimulation truly corresponds to moments of subjective urge relief.
    \item \textbf{Emotion and Memory Triggers:} In patients with implanted neurostimulators for depression or memory disorders, the wearable system can capture real-world mood fluctuations or memory recall episodes. Suppose an epilepsy patient with electrodes also has comorbid depression. The AI might detect periods of low activity/social isolation and sad affects (from voice and posture) and mark those, allowing study of depression-related brain states in daily life. Or if a patient listens to a particular song that evokes strong nostalgia, the system could record hippocampal activity to see how real autobiographical memory retrieval manifests in neural ensembles compared to lab word-list learning tasks.
\end{itemize}

In all these examples, the ability to pair the neural data with ground-truth context is transformative. We move beyond the clinic where a patient might press a button saying "I feel anxious now" – instead, we have sensor-verified context like "patient's heart rate spiked, breathing changed, and we detected they were crossing a busy intersection." This richness allows far better interpretation of neural signals.

Additionally, because the wearable AI is modular, researchers can program specific experiments within the real world. For instance, using the smart glasses, a researcher could occasionally present a standardized stimulus (like a quick cognitive test or an image) during daily life and immediately record the neural response – achieving an ambulatory version of an experiment, interleaved with life.

\subsection{Data and Technical Considerations}\label{ss:dataconsiderations}
Data collected in Research Mode can be quite large and complex. Each "event" yields a snippet of intracranial data (which could include multi-channel raw signals or even single-neuron spiking if microelectrodes are used) plus multimodal wearable data (video, audio, bio etc.). We anticipate using edge processing to extract key features (rather than saving hours of raw video, the LLM might generate summary descriptors like "location: kitchen, people: 2, activity: cooking" and log those alongside neural features like "theta power = X"). Still, for certain experiments, raw data might be needed and can be stored either on the local device (phones now have hundreds of GB storage) or uploaded to a secure cloud in bursts. Synchronization is achieved by time-stamping everything, and we can also align the implant's internal clock via telemetry.

One challenge is maintaining participant privacy, especially with continuous audio/video recording. In our envisioned implementation, raw video could be processed in real-time by the AI to look for predefined targets (like "detect if a person smiles" for a social study) and not stored or transmitted unless an event of interest is found. Also, participants can have options to pause data collection at any time for privacy. Ethically, this approach should be overseen by institutional review boards with careful consideration of data handling, much like current ambulatory EEG or bodycam research.

The benefit of this approach, however, is enormous. It helps close the measurement gap between human and animal neuroscience \citep{Mobbs2021}. In animals, researchers increasingly use wireless neural recording as the animal engages in naturalistic behavior, leading to discoveries of how neurons code complex actions or social hierarchy in group settings. Our system brings a similar capability to humans, who are arguably the most important subject of neuroscience but have been the most constrained in how we can study them. With tools like these and leveraging existing FDA approved implants, we foresee rapid progress in understanding neural mechanisms of human cognition and emotion in their normal habitat. This will enrich theoretical models and could also guide better clinical practices (as findings will be more ecologically valid).

Finally, it is worth pointing out the symbiosis with AI research: The data gathered—a massive set of aligned human behaviors and brain signals—could in turn be used to train better AI models that understand humans. Modern deep neural network architectures originated from early insights into biological neural processing but have since diverged considerably from their neurobiological roots. By collecting more extensive, real-world neural data, we can potentially realign these models with human-centric principles and draw fresh inspiration from actual brain organization and function. This aligns with ongoing efforts in brain-inspired and biologically grounded AI research, where insights from human neural circuits can inform next-generation architectures, driving more efficient, robust, and interpretable systems. In this sense, the research mode contributes not just to neuroscience but also to the advancement of AI, creating a virtuous cycle in which each field propels the other forward.

\section{Discussion}\label{s:discussion}

\subsection{Validation and Feasibility}\label{ss:validation}
Many components of our proposed system have been demonstrated in isolation. Closed-loop neural implants have shown therapeutic benefit in PTSD \citep{Gill2023} and other disorders. Wearable sensor algorithms can detect stress and PTSD events with reasonable accuracy \citep{Sadeghi2022}. Mobile intracranial recording technology (like Neuro-stack) has enabled patients to walk around while researchers record single neurons \citep{BRAINInitiative2023}. What remains is to integrate these into a unified platform. We suggest that integrating is now feasible given advances in mobile computing power and Multimodal LLM. A modern smartphone with on-board AI accelerators could run a compact multimodal transformer model analyzing streams from camera and biometrics in real time, especially if optimized (the tasks are akin to running a contextual assistant, which is within current tech limits and constantly upgraded). The RNS devices have low-latency detection (designed to respond to seizures quickly), which can be repurposed for our triggers. The communication between devices is a practical hurdle – e.g., current clinical RNS units don't stream live data due to battery limits – but next-generation implants are likely to have improved telemetry (for example, research devices like Medtronic Summit RC+S already allow streaming). Even in the interim, solutions like a smartphone tapping into an implant's data via NFC (an idea akin to using the iPhone as a magnet) can be tried. Nanthia Suthana's ongoing clinical trial work provides a testbed: her lab's PTSD trial (NCT04152993) could be extended by giving participants an early version of our wearable system to see if external triggers align with their internal data, potentially improving symptom tracking.

\subsection{Potential Challenges}\label{ss:challenges}
Interdisciplinary systems face challenges. One is user adherence – will patients consistently wear the gear? We envision using comfortable, discreet devices (smart glasses such as Meta Ray-Ban that look normal, a smart watch such as Apple Watch they'd wear anyway). As technology improves, sensors might integrate into everyday clothing (smart textiles, pendants, etc.). Privacy is a double-edged issue: patients might worry about being "watched" by AI, but on the other hand, those with severe PTSD might welcome the sense of a guardian system looking out for them. Ensuring all data is processed locally and giving the user control can mitigate privacy concerns. Another challenge is avoiding over-reliance on the system; the ultimate goal is to empower patients, not make them dependent on a crutch. That's why we emphasize using the system to train patients in non-invasive coping (the wearable loop) and ideally phasing out the implant.

\subsection{Interdisciplinary Collaboration}\label{ss:collaboration}
Building and validating this system requires collaboration across neuroscience, engineering, and AI disciplines. Neural engineers would refine the interfaces and ensure the implant and wearable can talk (ensuring electromagnetic safety, etc.). Neuroscientists and psychiatrists provide the domain knowledge on what signals and triggers matter (e.g., confirming amygdala theta as a biomarker, or what kind of messages calm a PTSD patient best). AI experts are needed to fine tune the multimodal LLM, efficiently edge-case deployment, and provide in context learning and alignment between model and human feedback. Ethicists and privacy experts must guide data governance. This kind of team science approach is increasingly recognized as necessary for next-gen neurotechnology \citep{SuthanaLab2025}. The BRAIN Initiative and similar efforts explicitly encourage integrating devices and data for human studies. Our proposal is a quintessential example – merging hardware, software, and clinical insight to achieve something that no single field could do alone.

\subsection{Comparison to Other Approaches}\label{ss:comparison}
Traditional PTSD therapies like prolonged exposure or cognitive processing therapy aim to gradually desensitize patients to triggers by recalling them in safe settings. Our system in a way automates and extends exposure therapy into everyday life: the AR cues could gently expose patients to mild triggers and then calm them, potentially accelerating habituation. Pharmacologically, drugs like prazosin are used for nightmares, and beta-blockers for situational anxiety; these are generalized approaches, whereas our system is personalized and adaptive in the context. It may also have fewer side effects than systemic medication. Other devices like vagus nerve stimulators or transcranial magnetic stimulators have been tried for PTSD, but those are open-loop and lack the contextual awareness our system has. By being responsive and context-integrated, our approach aims to maximize efficacy and minimize unnecessary intervention.

On the research front, our approach stands out by collecting data continuously in natural life. Functional MRI or MEG provides great spatial mapping but requires the subject to lie still in a lab; our data is complementary, focusing on dynamics and behavior. Recent wearable brain imaging like on-scalp MEG helmets are promising \citep{Trust2018}, but those measure from outside the skull and have lower signal fidelity than direct electrodes – they could, however, be combined with our platform for less invasive monitoring if resolution improves \citep{Trust2018}. What really distinguishes our system is the Multimodal LLM-driven event detection and super-human level contextual understanding, enabling capture of specific moments of interest; this reduces data overload and highlights salient periods for analysis.

\subsection{Toward Revolutionizing Neuroscience with AI}\label{ss:revolution}
Looking forward, the implications of ubiquitous AI-enabled neurotechnology are profound. We can start to envision a world where brain monitoring and mental health support are ever-present but unobtrusive, similar to how fitness trackers normalized constant physical health monitoring. For chronic psychiatric conditions prone to relapse (PTSD, OCD, depression, etc.), having a 24/7 guardian system could dramatically improve long-term outcomes by ensuring early detection of issues and timely intervention – preventing small fires from becoming wildfires, so to speak. The data harvested would allow clinicians to move to data-driven psychiatry, where diagnoses and treatments are informed by objective patterns (for example, identifying a subtype of PTSD that has a distinct physiological trigger signature and tailoring therapy accordingly).

From a neuroscience perspective, we would gain the ability to test hypotheses about brain function under conditions that truly matter to people's lives. As Mobbs et al. (2021) articulate, applying computational ethology methods to humans can reveal behavioral motifs and neural dynamics that are hidden in constrained lab tasks \citep{Mobbs2021}. Our system is a concrete step toward that vision. It narrows the gap between human and animal research, allowing paradigms like free-roaming foraging (going about town) or natural social hierarchies (family dinner) to be studied at the neural level in humans. The result could be new discoveries about memory encoding in real scenes, emotional regulation during actual stress, and how neural circuits multitask in complex environments – insights that could not only advance basic science but also inform more naturalistic forms of AI.

\section{Conclusion}\label{s:conclusion}
We have outlined a comprehensive dual-loop system that marries cutting-edge neurotechnology with artificial intelligence to both treat PTSD and propel neuroscience research into the real world. In PTSD therapy, the combination of an implanted RNS device watching over the brain's emotional network and an LLM-driven wearable ensemble watching over the patient's environment represents a new paradigm of care – one that is proactive, personalized, and context-aware. Early clinical evidence supports the feasibility: abnormal deep-brain signals can be detected and modulated for therapeutic gain \citep{Gill2023}, and external triggers can be sensed and managed by wearable technology \citep{Sadeghi2022,NightWare}. Our contribution is to integrate these into a cohesive closed-loop that shields the patient on both fronts, inner and outer, and crucially, learns from each interaction to become smarter and less invasive over time.

In the domain of neuroscience research, by equipping participants with AI that synchronizes brain recordings with lived experience, we break the shackles of the laboratory. Complex human behaviors and social-emotional experiences – once thought too unwieldy to study with neural precision – can now be probed. This system essentially turns the world into a laboratory while preserving its authenticity. The implications range from improving our understanding of disorders like PTSD (by seeing how the brain and body respond in the wild) to fundamental discoveries about human cognition.

Realizing this vision will require cross-pollination between fields. Neuroscientists must embrace AI as a tool for parsing behavioral context, and AI researchers must embrace the messiness of real-world human data as a frontier for next generation model development. The mutual benefits cannot be overstated: neuroscience provides grounding and purpose for AI (what could be more meaningful than alleviating human suffering and uncovering the workings of our own minds?), and AI provides neuroscience with powerful new capabilities to make sense of high-dimensional data and control complex systems. This symbiosis is emblematic of the emerging era of "hybrid intelligence" where humans and AI cooperate continuously. Our dual-loop system is an early exemplar – a partnership between a human nervous system and an AI assistant, mediated by existing technology feasible and regulatory approved devices, working together to maintain health and drive discovery.

In closing, we envision a future in which AI-powered multimodal neurotechnology revolutionizes mental health care and brain science. Patients with PTSD (and other conditions) could have round-the-clock support, experiencing a level of safety and empowerment previously unattainable, as if they carry a personal therapist and neurologist with them at all times. Meanwhile, every moment of their journey contributes data to a growing understanding of the human brain in its natural habitat, bridging the gap between lab theories and life's realities. By freeing neuroscience from the lab, we allow it to truly see and heal the brain in the context of the whole person. This future is ambitious but within reach, and pursuing it stands to benefit countless individuals and enrich our knowledge of ourselves. It is an opportunity and arguably a responsibility for the neuroscience and AI communities to collaborate to leverage technology for the betterment of human life.

\bibliography{bibliography}

\begin{thebibliography}{17}
\providecommand{\natexlab}[1]{#1}
\providecommand{\url}[1]{\texttt{#1}}
\expandafter\ifx\csname urlstyle\endcsname\relax
  \providecommand{\doi}[1]{doi: #1}\else
  \providecommand{\doi}{doi: \begingroup \urlstyle{rm}\Url}\fi

\bibitem[AlSaad et~al.(2024)AlSaad, Abd-Alrazaq, Boughorbel, Ahmed, Renault, Damseh, and Sheikh]{AlSaad2024}
Rawan AlSaad, Alaa Abd-Alrazaq, Sabri Boughorbel, A.~Ahmed, M.-A. Renault, Rafat Damseh, and Javaid Sheikh.
\newblock Multimodal large language models in healthcare: Applications, challenges, and future outlook (preprint).
\newblock \emph{Journal of Medical Internet Research}, 26:\penalty0 e59505, 2024.
\newblock \doi{10.2196/59505}.

\bibitem[{BRAIN Initiative}(2023)]{BRAINInitiative2023}
{BRAIN Initiative}.
\newblock Brain researchers develop a wearable system to record brain activity in freely moving humans, mar 2023.
\newblock URL \url{https://braininitiative.nih.gov/news-events/blog/brain-researchers-develop-wearable-system-record-brain-activity-freely-moving}.

\bibitem[{Daily Bruin}(2023)]{DailyBruin2023}
{Daily Bruin}.
\newblock Neuro-stack device technology shows promise for future of brain research, apr 2023.
\newblock URL \url{https://dailybruin.com/2023/04/03/neuro-stack-device-technology-shows-promise-for-future-of-brain-research}.

\bibitem[Englot et~al.(2017)Englot, Birk, and Chang]{Englot2017}
Dario~J. Englot, Harjus Birk, and Edward~F. Chang.
\newblock Seizure outcomes in non-resective epilepsy surgery: An update.
\newblock \emph{Neurosurgical Review}, 40\penalty0 (2):\penalty0 181--194, 2017.
\newblock \doi{10.1007/s10143-016-0725-8}.

\bibitem[Gill et~al.(2023)Gill, Schneiders, Stangl, Aghajan, Vallejo, Hiller, Topalovic, Inman, Villaroman, Bari, Adhikari, Rao, Fanselow, Craske, Krahl, Chen, Vick, Hasulak, Kao, and Koek]{Gill2023}
Jaya~L. Gill, J.~Anthony Schneiders, Matthias Stangl, Zahra~M. Aghajan, Manuel Vallejo, Sebastian Hiller, Uros Topalovic, Cory~S. Inman, Daniel Villaroman, Ausaf Bari, Avishek Adhikari, Vikram~R. Rao, Michael~S. Fanselow, Michelle~G. Craske, Scott~E. Krahl, Jean W.~Y. Chen, Mark Vick, Nancy~R. Hasulak, Jonathan~C. Kao, and Ralph~J. Koek.
\newblock A pilot study of closed-loop neuromodulation for treatment-resistant post-traumatic stress disorder.
\newblock \emph{Nature Communications}, 14\penalty0 (1):\penalty0 2997, 2023.
\newblock \doi{10.1038/s41467-023-38712-1}.

\bibitem[Javanbakht et~al.(2024)Javanbakht, Hinchey, Gorski, Ballard, Ritchie, and Amirsadri]{Javanbakht2024}
Arash Javanbakht, Lisa Hinchey, Katherine Gorski, Amber Ballard, Lisa Ritchie, and Alireza Amirsadri.
\newblock Unreal that feels real: Artificial intelligence-enhanced augmented reality for treating social and occupational dysfunction in post-traumatic stress disorder and anxiety disorders.
\newblock \emph{European Journal of Psychotraumatology}, 15\penalty0 (1), 2024.
\newblock \doi{10.1080/20008066.2024.2418248}.

\bibitem[Kalisperakis et~al.(2023)Kalisperakis, Karantinos, Lazaridi, Garyfalli, Filntisis, Zlatintsi, Efthymiou, Mantas, Mantonakis, Mougiakos, Maglogiannis, Tsanakas, Maragos, and Smyrnis]{Kalisperakis2023}
Emmanouil Kalisperakis, T.~Karantinos, M.~Lazaridi, V.~Garyfalli, P.~P. Filntisis, Athanasia Zlatintsi, N.~Efthymiou, A.~Mantas, Leonidas Mantonakis, Theodoros Mougiakos, Ilias Maglogiannis, Panayotis Tsanakas, P.~Maragos, and Nikolaos Smyrnis.
\newblock Smartwatch digital phenotypes predict positive and negative symptom variation in a longitudinal monitoring study of patients with psychotic disorders.
\newblock \emph{Frontiers in Psychiatry}, 14, 2023.
\newblock \doi{10.3389/fpsyt.2023.1024965}.

\bibitem[Kellogg et~al.(2024)Kellogg, Ernst, Spencer, Datta, Klein, Bhati, Shivacharan, Nho, Barbosa, Halpern, and Raslan]{Kellogg2024}
Marissa~A. Kellogg, Laura~D. Ernst, David~C. Spencer, Proleta Datta, Eoin Klein, Mahendra~T. Bhati, Rajat~S. Shivacharan, Young-Hoon Nho, Daniel A.~N. Barbosa, Casey~H. Halpern, and Ahmed Raslan.
\newblock Dual treatment of refractory focal epilepsy and obsessive-compulsive disorder with intracranial responsive neurostimulation.
\newblock \emph{Neurology Clinical Practice}, 14\penalty0 (4), 2024.
\newblock \doi{10.1212/cpj.0000000000200318}.

\bibitem[Mobbs et~al.(2021)Mobbs, Wise, Suthana, Guzmán, Kriegeskorte, and Leibo]{Mobbs2021}
Dean Mobbs, Toby Wise, Nanthia Suthana, Natalia Guzmán, Nikolaus Kriegeskorte, and Joel~Z. Leibo.
\newblock Promises and challenges of human computational ethology.
\newblock \emph{Neuron}, 109\penalty0 (14):\penalty0 2224--2238, 2021.
\newblock \doi{10.1016/j.neuron.2021.05.021}.

\bibitem[Nho et~al.(2023)Nho, Rolle, Topalovic, Shivacharan, Cunningham, Hiller, Batista, Feng, Espil, Kratter, Bhati, Kellogg, Raslan, Williams, Garnett, Pesaran, Oathes, Suthana, Daniel, and Halpern]{Nho2023}
Young-Hoon Nho, Camarin~E. Rolle, Uros Topalovic, Rajat~S. Shivacharan, Trevor Cunningham, Sebastian Hiller, Daniel~Macêdo Batista, Austin~Y. Feng, Flint~M. Espil, Ian~H. Kratter, Mahendra~T. Bhati, Marissa Kellogg, Ahmed~M. Raslan, Nolan Williams, James Garnett, Bijan Pesaran, Desmond~J. Oathes, Nanthia Suthana, Daniel, and Casey~H. Halpern.
\newblock Responsive deep brain stimulation guided by ventral striatal electrophysiology of obsession durably ameliorates compulsion.
\newblock \emph{Neuron}, 2023.
\newblock \doi{10.1016/j.neuron.2023.09.034}.

\bibitem[{NightWare}(n.d.)]{NightWare}
{NightWare}.
\newblock Product, n.d.
\newblock URL \url{https://nightware.com/product/}.

\bibitem[Sadeghi et~al.(2022)Sadeghi, McDonald, and Sasangohar]{Sadeghi2022}
Mahnoosh Sadeghi, Anthony~D. McDonald, and Farzan Sasangohar.
\newblock Posttraumatic stress disorder hyperarousal event detection using smartwatch physiological and activity data.
\newblock \emph{PLOS ONE}, 17\penalty0 (5):\penalty0 e0267749, 2022.
\newblock \doi{10.1371/journal.pone.0267749}.

\bibitem[Stangl et~al.(2023)Stangl, Maoz, and Suthana]{Stangl2023}
Matthias Stangl, Sonya~L. Maoz, and Nanthia Suthana.
\newblock Mobile cognition: Imaging the human brain in the "real world", jun 2023.
\newblock URL \url{https://dukespace.lib.duke.edu/items/a9193bcc-246f-4655-85e8-be160f2addd6}.

\bibitem[{Suthana Lab}(2025)]{SuthanaLab2025}
{Suthana Lab}.
\newblock Ptsd clinical trial and research study, 2025.
\newblock URL \url{https://suthanalab.com/ptsd/}.

\bibitem[{Texas A\&M Engineering}(2017)]{TexasAMEngineering2017}
{Texas A\&M Engineering}.
\newblock Wearable ptsd monitor one step closer to reality, jun 2017.
\newblock URL \url{https://tees.tamu.edu/news/2017/06/wearable-ptsd-monitor-one-step-closer-to-reality.html}.

\bibitem[Trust(2018)]{Trust2018}
W.~Trust.
\newblock New wearable brain scanner allows patients to move freely for the first time, March~21 2018.
\newblock URL \url{https://medicalxpress.com/news/2018-03-wearable-brain-scanner-patients-freely.html}.

\bibitem[Wang \& Wen(2025)Wang and Wen]{Wang2025}
Edward~Hong Wang and Cynthia~Xin Wen.
\newblock Building a unified ai-centric language system: analysis, framework and future work.
\newblock \emph{arXiv preprint arXiv:2502.04488}, Feb 2025.
\newblock URL \url{https://arxiv.org/abs/2502.04488}.
\newblock arXiv:2502.04488v1 [cs.CL].

\end{thebibliography}
\bibliographystyle{bibliography}

\end{document}